\documentclass[journal,comsoc]{IEEEtran}
\usepackage[T1]{fontenc}

\ifCLASSINFOpdf
\else
\fi
\usepackage{amsmath}
\usepackage[cmintegrals]{newtxmath}
\usepackage[utf8]{inputenc}
\usepackage{graphicx}
\usepackage[justification=centering,labelfont=bf]{caption}
\usepackage[english]{babel}
\usepackage{amsmath,amssymb,amsfonts}
\usepackage{algorithm,algorithmic}
\usepackage{graphicx}
\usepackage[breakable]{tcolorbox}
\usepackage{textcomp}
\usepackage{nicefrac}
\usepackage{mathtools}
\usepackage[short]{optidef}
\usepackage{ifthen}
\usepackage{booktabs}
\usepackage{multirow}
\usepackage{makecell}
\usepackage{wrapfig}
\usepackage{scalerel}
\usepackage[makeroom]{cancel}
\usepackage{fixmath}
\usepackage{enumitem}

\newcommand{\vast}{\bBigg@{4}}
\newcommand{\Vast}{\bBigg@{5}}

\newlength\bshft
\def\fakebold#1{\ThisStyle{\ooalign{$\SavedStyle#1$\cr%
			\kern-\bshft$\SavedStyle#1$\cr%
			\kern\bshft$\SavedStyle#1$}}}

\newcommand{\I}{\fakebold{\mathbb{I}}}

\newcommand{\R}{\mathbb{R}}

\newcommand{\C}{\mathbb{C}}

\newcommand{\defeqto}{\vcentcolon  =}
\renewcommand{\vec}{\mathbold}
\renewcommand{\j}{\jmath\,}
\newcommand{\inc}{\mathrm{inc}}

\newcommand{\obs}{\mathrm{obs}}
\newcommand{\vo}{\mathcal{E}}

\newcommand{\Ps}{\mathcal{P}}
\newcommand{\e}{\mathrm{e}}
\newcommand{\diag}{\mathrm{diag}}

\newcommand{\cur}{I}
\newcommand{\tx}{\mathrm{tx}}
\newcommand{\rx}{\mathrm{rx}}
\newcommand{\relem}{r^\mathrm{elem}_\textsc{ff}}
\newcommand{\rarray}{r_\textsc{ff}}
\newcommand{\am}{\vec{\tilde{a}}}

\begin{document}
%
\title{Reconfigurable Intelligent Surfaces:\\Bridging the gap between scattering and reflection}
%
%
%

\author{Juan Carlos~Bucheli Garcia,~\IEEEmembership{Student Member,~IEEE,}\thanks{J. Bucheli is pursuing his doctoral degree with Huawei Technologies at the Mathematical and Algorithmic Sciences Lab. Boulogne-Billancourt, France. and Telecom Paris. Institut Polytechnique de Paris. Palaiseau, France. (e-mail: juanc.garcia1@huawei.com, juan.bucheli-garcia@telecom-paris.fr).}
Alain~Sibille,~\IEEEmembership{Senior~Member,~IEEE,}\thanks{A. Sibille is with the \textit{Laboratoire Traitement et Communication de l'Information}, LTCI, Telecom Paris. Institut Polytechnique de Paris. Palaiseau, France. (e-mail: alain.sibille@telecom-paris.fr).}
and~Mohamed~Kamoun,~\IEEEmembership{Member,~IEEE.}\thanks{M. Kamoun is with Huawei Technologies at the Mathematical and Algorithmic Sciences Lab. Boulogne-Billancourt, France. (e-mail: \mbox{mohamed.kamoun@huawei.com}).}
}
\maketitle

\begin{abstract}
In this work we address the distance dependence of reconfigurable intelligent surfaces (RIS). As differentiating factor to other works in the literature, we focus on the array near-field, what allows us to comprehend and expose the promising potential of RIS. The latter mostly implies an interplay between the physical size of the RIS and the size of the Fresnel zones at the RIS location, highlighting the major role of the phase. 

To be specific, the point-like (or zero-dimensional) conventional scattering characterization results in the well-known dependence with the fourth power of the distance. On the contrary, the characterization of its near-field region exposes a reflective behavior following a dependence with the second and third power of distance, respectively, for a two-dimensional (planar) and one-dimensional (linear) RIS. Furthermore, a smart RIS implementing an optimized phase control can result in a power exponent of four that, paradoxically, outperforms free-space propagation when operated in its near-field vicinity. All these features have a major impact on the practical applicability of the RIS concept.

As one contribution of this work, the article concludes by presenting a complete signal characterization for a wireless link in the presence of RIS on all such regions of operation.




\end{abstract}


%
\IEEEpeerreviewmaketitle

\section{Introduction}

The enhancement of wireless connectivity in the last decades has radically changed the way we humans perceive and interact with our surroundings. Nonetheless, the interfacing network infrastructure has been mostly confined at rooftops and distant-away serving sites. 

Recently, the ancient idea of improving wireless networks by means of relays has been renovated through the concept of low-cost smart mirrors. As a consequence, nowadays' scientific literature is full of appellatives such as intelligent reflecting surfaces (IRS), large intelligent surfaces (LIS), reconfigurable intelligent surfaces (RIS), passive relaying arrays (PRA), among others~\cite{RIS,LIS,IRS,WCNC2019}. 

The underneath idea behind these is to control the characteristics of the radio channel. Even-though such an idea is in a sense revolutionary, most of the scientific literature is concentrated on algorithmic and signal processing aspects. In fact, the comprehension of the involved electromagnetic specificities has not been fully addressed.

More specifically, \textit{reconfigurable intelligent surfaces} (RIS) correspond to the arrangement of a massive amount of inexpensive antenna elements with the objective of capturing and scattering energy in a controllable manner. Such a control method varies widely in the literature~\cite{RIS}; among which PIN-diode and varactor based are popular. 

In this context, the authors investigate an impedance controlled RIS, although the addressed fundamentals are of a much wider applicability. As a matter of fact, by characterizing RIS in terms of the elements' observed impedance, it is possible to study multiple variants. Nonetheless, with energy efficiency considerations in mind, the current work will focus on the passive and non-dissipative purely reactive alternative. 

It is well known from the radar community that, while a mirror is a large reflecting surface with its reflected energy decreasing with the 2\textsuperscript{nd} power of distance, a scatterer is usually considered a near-point object with the scattered energy falling with the 4\textsuperscript{th} power of the distance~\cite{BOOKBALANIS}. 

As a matter of fact, to the authors knowledge, there is no consolidated concern in the literature about the nature of the RIS as a scatterer or as a mirror. In fact, although most works seem to use both terms interchangeably, the conventional view has been through their characterization as scatterers.

Particularly, the authors of~\cite{RIS} departed from a generalization of the two-ray channel model to argue that the RIS does not necessarily obey a path-loss dependence with the fourth power of distance. On the other hand, the authors of~\cite{IRS1} concluded there that such a strong power law is probably unavoidable for a practical RIS. Nevertheless, no reference to the crucial role of the array near-field region was found as a way to explain such discrepancies.

Consequently, in this work, we offer a view that unifies the previous seemingly opposite scattering/reflection dual perspectives as a mean to identify scenarios and show the strong potential behind the RIS concept. In particular, we show how physical area and distance aspects become of paramount importance for the operation of such devices.

The rest of the paper is organized as follows: the next section begins by deepening on the importance of the near-field region for RIS and, subsequently, section~\ref{s:prop} studies the implications of operating it over such a region. 

In particular, the latter is based on the derived mathematical model and, as well, on the understanding of the interacting Fresnel zones. Additionally, the model and established notions are verified through simulation results obtained from a commercial electromagnetic solver. 

Further, section~\ref{s:RISBTFS} explains how RIS offers the possibility of outperforming free-space propagation over completely obstructed direct transmitter-receiver links. 

Finally, the authors conclude by presenting in section~\ref{s:llmodel} a complete signal-level characterization for a transmitter-receiver link in the presence of RIS, which captures all the described phenomena. 
\section{The RIS far-field decomposition}
\label{s:ffdecom}
Let us begin by recalling that the far-field approximation imposes a minimum transmitter-receiver separation distance, notably, so that conventional antenna and propagation models are valid. In fact, the far-field distance increases with the square of the largest dimension of the antenna. 

To be clear, in the standard cell-centric network architecture, the far-field approximation has greatly sufficed as means of characterization. Nevertheless, for the case of RIS, one large issue at stake can be stated as a paradox and, also, related to the fact that it has been mostly conceived as an entirely passive\footnote{-- in the sense that it does not inherently inject energy to the environment.} architecture.

Specifically, a RIS must be large as a mean of capturing enough energy; but as it grows, conventional far-field decomposition mandates that the transmitter and receiver must move away. Consequently, the larger the RIS gets, the farther the transmitter and receiver must be and, therefore, the stronger the path-loss of the transmitter-RIS-receiver link.

Fortunately, this paradox can be circumvented by operating RIS over its near-field\footnote{It must be stressed that we are referring exclusively to the near-field of the array and not to the near-field of the array elementary unit itself.} region. In order to understand the relevance of its near-field, consider the RIS essentially as the arrangement of multiple antenna elements.

\begin{figure}[t]
	\centerline{\includegraphics[width=0.45\textwidth]{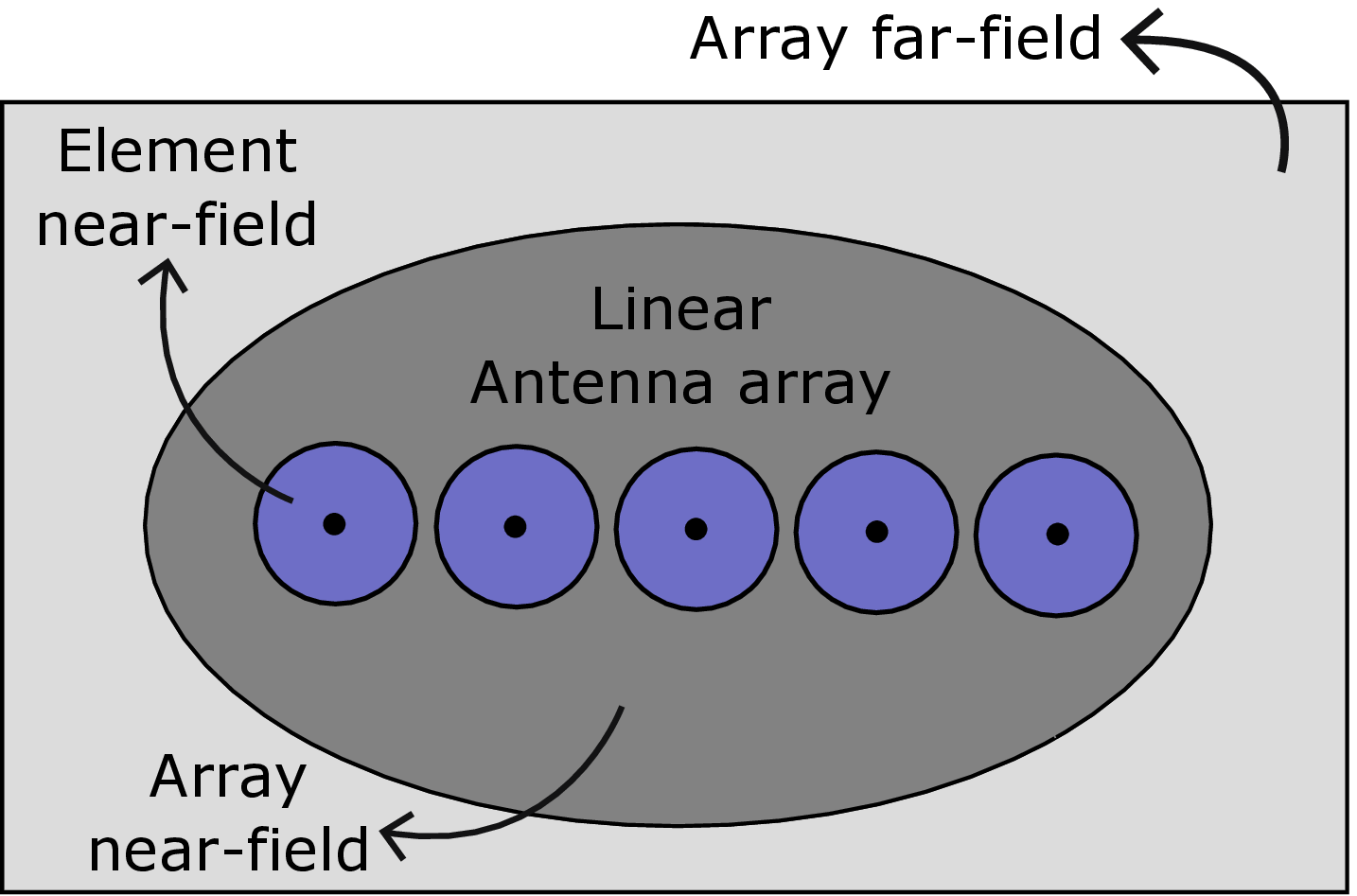}}
	\caption{Pictorial representation of the field regions of a linear antenna arrangement.}
	\label{fig:venn1}
\end{figure}

More specifically, Fig.~\ref{fig:venn1} shows explicitly the array far-field and the array near-field as disjoint regions. The element near-field is also presented in dark blue for reference. Nonetheless, the element near-field is not considered a region of interest in this work as it is generally too close to the RIS.

As a matter of fact, any array can be approached close enough to be in the far-field of each elementary unit but not in the far-field of the full array. The authors of this work refer to that particular region of space as the array near-field.


The importance of the array near-field characterization is that a RIS requires large antenna arrangements in the close proximity of users, unlike traditional transmitter-receiver links. Therefore, the term \textit{array near-field} is extensively used here as a way to differentiate it, given its importance in the case of large enough RIS. 

In the array far-field, the directional characteristics of RIS naturally decouple from the separation distance. On the other hand, in the array near-field, the previous dependencies are more intricate and call for a different understanding of the problem.

For a simpler insight and self-consistency, appendix~\ref{ap:EJ} and~\ref{ap:EJqff} are devoted to the fundamental derivations of the far-field and array near-field based on elementary Maxwell's equations. To be specific, Appendix~\ref{ap:EJ} reviews the conventional solution for far-field radiation problems, in order to familiarize with the adopted notation. Subsequently, Appendix~\ref{ap:EJqff} formally introduces the array near-field characterization via the here-called generalized array manifold.

\section{RIS-mediated propagation}
\label{s:prop}

In light of the difficulties of the conventional far-field view to capture the real scenarios of interest, the current section takes a look at how RIS can be characterized using the near-field metrics described in Appendix~\ref{ap:EJqff}.

More specifically, the section begins by reviewing existing work on modeling of a specific RIS-like architecture in its far-field. Subsequently, relying on the generalized array manifold, such model is extended to RIS over the array near-field region. 

Altogether, the derived model is subsequently studied on its canonical configuration\footnote{-- the one in which the RIS elements are short-circuited.} in order to demonstrate its validity based on results obtained from a commercially-available electromagnetic solver. 

In what follows, we also expose how the Fresnel zone decomposition at the RIS interface can provide a powerful and intuitive understanding of its promising potential.

\subsection{RIS operating in the array far-field region}
\label{s:RISGff}

The authors of this paper derived in~\cite{WCNC2019} the model for a dipole-based Passive Relaying Array (PRA). Particularly, a PRA can be regarded as a bulky RIS in which the elements are not necessarily disposed conformal to a surface. Likewise, the control of PRA is assumed to be mediated through dynamically tunable reactive loads.

Let us consider the derived model for the \(\theta\)-polarized scattered field in the array far-field region of the array:
\begin{equation}
\label{eq:EScat}
E_\theta(\hat{\vec{r}}_\inc,\hat{\vec{r}}_\obs) = \j\frac{\eta\,\e^{-\j kr}}{k \pi r}\am^H(\hat{\vec{r}}_\obs)\big(\vec{Z}+\j\vec{X}_L\big)^{-1}\,\am(\hat{\vec{r}}_\inc)\,E_{\inc,\theta},
\end{equation}
where \(\vec{Z}\in\C^{N{\times}N}\) is the impedance matrix~\cite{BOOKBALANIS} of such an \(N\) element array; \(\vec{X}_L=\diag(\vec{x})\in\R^{N{\times}N}\) is a diagonal matrix containing the values of the controllable reactive loads; \(r\) is the PRA-observer distance; \(\am(\hat{\vec{r}})\) is the there-called modified steering vector\footnote{The \textit{modified} denomination was added given that, unlike the conventional steering vector, the one in~\cite{WCNC2019} includes the impact of the element pattern.}; and \(\hat{\vec{r}}_\inc\) and \(\hat{\vec{r}}_\obs\) are two (outward) unitary vectors pointing towards the source and observer, respectively.

In spite of PRA's structural assumptions, the model derived in~\cite{WCNC2019} is valid for architectures with arbitrarily disposed elements. 

Moreover, in the specific case of such a PRA, the architecture was based on dipoles as they are minimum scattering antennas (MSA)~\cite{MSA}, allowing to entirely characterize its scattering behavior by the aid of the antenna circuit representation. 

Nonetheless, in the case of arbitrary array elementary units, the missing piece is utterly the contribution of the structural mode~\cite{ASMODES}, which can be computed once and for all as it is constant with respect to the controllable loads (that are commonly assumed to be the available degrees of freedom).

Note that, in the derivation of~\eqref{eq:EScat}, the impact of the transmitting antenna pattern was assumed negligible as we were assuming it to be in the array far-field. 

In what follows, and for simplicity of presentation, we will resort to the assumption that the transmitter antenna is an isotropic source of fields. 

\subsection{The field scattered by RIS at the array near-field region}
\label{s:RISQff}

To begin with, it must be stressed that the array far-field approximation of both the reception and transmission processes\footnote{-- those which jointly compose the scattering process.} is implicitly captured in~\eqref{eq:EScat} by the modified steering vector \(\am(\vec{\hat{r}})\) and, as well, by the distance dependence. 


Nonetheless, as exposed in Appendix~\ref{ap:EJqff}, an expression valid for the array near-field can be obtained by identifying the steering vector as a special case of the generalized array manifold, the latter of which is given by:
\begin{equation}
\label{eq:AMv}
\vec{a}_p(\vec{r})\big|_n \defeqto G(\vec{r}-\vec{r}_n)\,F_{0,p}\bigg(\frac{\vec{r}-\vec{r}_n}{|\vec{r}-\vec{r}_n|}\bigg)\,\forall\,n\le N,
\end{equation}
with \(G(\vec{r})\) being the free-space Green function of Appendix~\ref{ap:EJ} and \(F_{0,p}(\vec{\hat{r}})\) the radiation vector of the array elementary unit along the p direction of polarization.

\begin{figure}[t]
	\centerline{\includegraphics[width=0.45\textwidth]{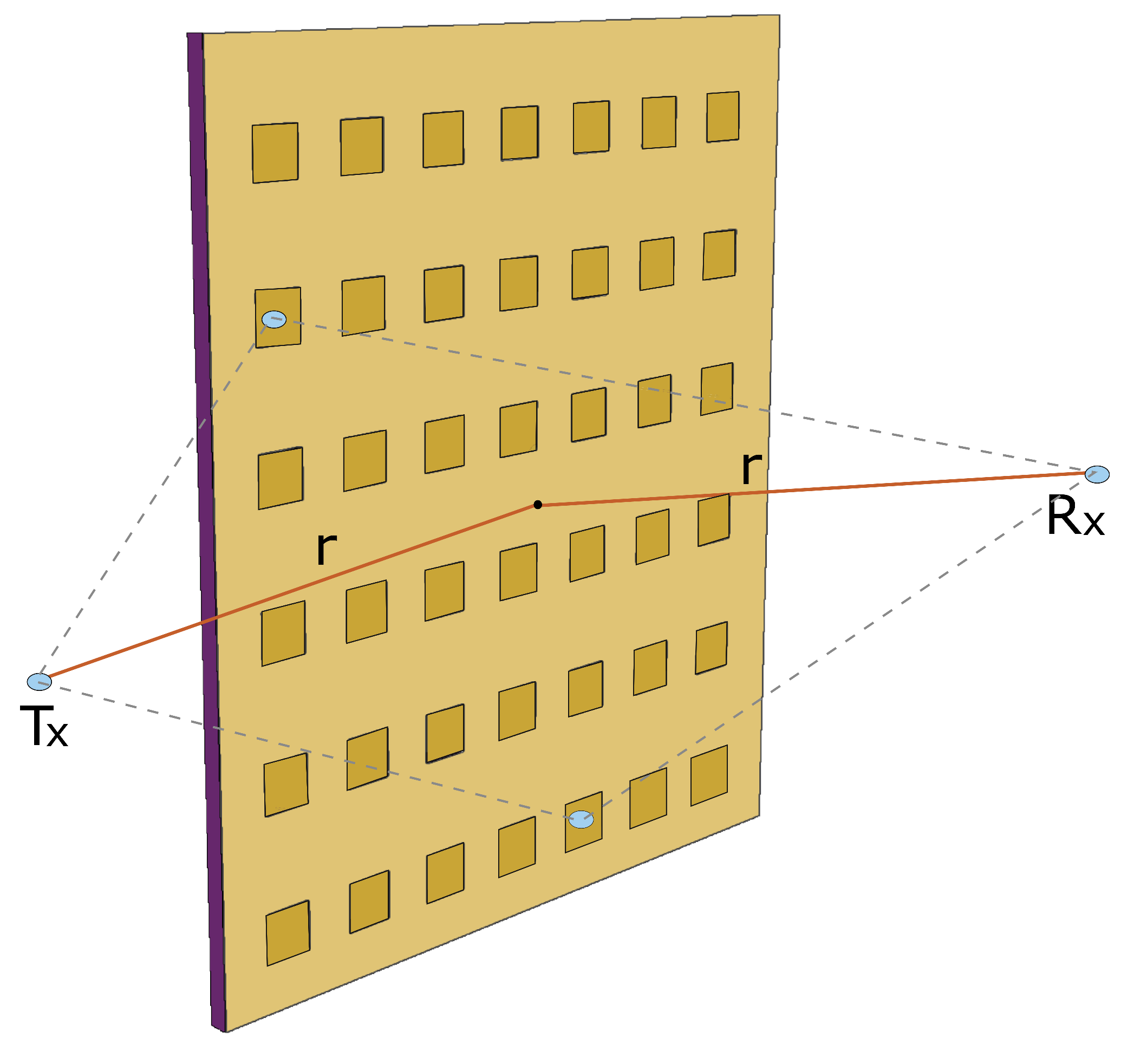}}
	\caption{Schematic view of Tx, Rx and RIS.}
	\label{fig:planarRIS}
\end{figure}

By doing so, it can be shown that the scattered electric field intensity at the array near-field reads:
\begin{equation}
\label{eq:ERIS}
	E_\theta(\vec{r}_\rx,\vec{r}_\tx) = k^2\eta^2\,\vec{a}_\theta^H(\vec{r}_\rx)\big(\vec{Z}+\j\vec{X}_L\big)^{-1}\,\vec{a}_\theta(\vec{r}_\tx)\,F_{\theta}^\textrm{iso}\,I_\tx,
\end{equation}
where \(\vec{r}_\tx\) and \(\vec{r}_\rx\) are the complete coordinates\footnote{-- as opposed to~\eqref{eq:EScat} where \(|\vec{r}_\tx|>\rarray\) and \(|\vec{r}_\rx|>\rarray\) and, therefore, there was only a dependence on the directions of incidence and observation.} of the transmitter and receiver relative to RIS' coordinate reference (e.g. the RIS center as in Fig.~\ref{fig:planarRIS}) and, as we are dealing with \(\vec{\hat{z}}\)-oriented dipoles, only the \(\theta\) polarization is considered. 

Moreover, note that the dependence on the input current to the transmitter antenna \(I_\tx\) exposes the role of its respective radiation vector (i.e. \(F_{\theta}^\textrm{iso}\)).

Observe also that \(\big(\vec{Z}+\j\vec{X}_L\big)^{-1}\) is a transpose symmetric matrix, which has often (e.g.~\cite{RIS,LIS,IRS}) been characterized as a diagonal matrix containing complex exponential factors that account for digitally-tunable phase shifts.

For simplicity, let us disregard the phenomenon of mutual coupling (i.e. \(\vec{Z} = Z_A\,\I_{N{\times}N}\)) and, additionally, short circuit all elements (\(\vec{x}=\vec{0}\,\Omega\)). Under those circumstances,~\eqref{eq:ERIS} can be simply expressed as:
\begin{equation}
\label{eq:ERISUnc}
	E_\theta(\vec{r}_\rx,\vec{r}_\tx) = \frac{k^2\eta^2}{Z_A}\vec{a}_\theta^H(\vec{r}_\rx)\,\vec{a}_\theta(\vec{r}_\tx)\,F_{\theta}^\textrm{iso}\,I_\tx.
\end{equation}

If, additionally, we assume that the RIS is provided with \(\theta\)-polarized isotropic elementary units, i.e. \(\vec{F}_{\perp}(\vec{\hat{r}})=\vec{\hat{\theta}}F_{\theta}^\textrm{iso}\), the expression~\eqref{eq:ERISUnc} can be simplified further to:
\begin{equation}
\label{eq:ERISUncIso}
E_\theta(\vec{r}_\rx,\vec{r}_\tx) = \frac{k^2\eta^2}{Z_A}\big[F_{\theta}^\textrm{iso}\big]^3 I_\tx \sum\limits_{n=1}^N G\big(\vec{r}_\rx-\vec{r}_n\big)\,G\big(\vec{r}_\tx-\vec{r}_n\big),
\end{equation}
where \(G(\vec{r})\) is, once more, the Green function of Appendix~\ref{ap:EJ}.

Even-though the expression in~\eqref{eq:ERISUncIso} corresponds to the array near-field electric field intensity, it characterizes the linear combination of (locally) far-field sources. Thus, the radiation density (power per unit area) resulting from the RIS can be related simply to its scattered field through \(\Ps=(2\eta)^{-1}|\vec{E}|^2\). 

In particular, introducing \(r_{\textrm{t},n}{\defeqto}|\vec{r}_\tx-\vec{r}_n|\) and \(r_{\textrm{r},n}{\defeqto}|\vec{r}_\rx-\vec{r}_n|\) as the distances from the transmitter and receiver to every RIS element, respectively, the radiation density of the scattered field reads: 
\begin{equation}
\label{eq:RDISUncIso}
	\Ps(\vec{r}_\rx,\vec{r}_\tx) = \frac{k^4\eta^3}{2|Z_A|^2}\big|F_{\theta}^\textrm{iso}\big|^6 |I_\tx|^2 \Bigg|\sum\limits_{n=1}^N \frac{\e^{-\j kr_{\textrm{r},n}}}{4\pi r_{\textrm{r},n}}\,\frac{\e^{-\j kr_{\textrm{t},n}}}{4\pi r_{\textrm{t},n}}\Bigg|^2.
\end{equation}

Observe that, even-though we have assumed hypothetical isotropic elements,~\eqref{eq:RDISUncIso} allows to analyze the radiation density as regards the transmitter and receiver locations relative to the element's disposition, i.e. \(\vec{r}_n\,\forall\,n\le N\).

In the scenario of Fig.~\ref{fig:planarRIS} is shown a symmetrical setup in which the transmitter and receiver are both at a distance \(r\) from the center of the RIS, which is composed of a 2D periodic structure along its planar surface with a total of \(N\) elements.

Moreover, in light of the loading condition under evaluation (\(\vec{x}=\vec{0}\,\Omega\)) in~\eqref{eq:RDISUncIso}, the transmitter and receiver are positioned symmetrically (\(45^\circ\) from the vector normal to RIS' surface) as required by the Snell-Decartes law of reflection. The latter, in order to expose a case in which waves are naturally interfering constructively\footnote{In fact, they don't interfere perfectly (as it will be clear later) but such a setup serves to illustrate the point the authors want to make.} towards the receiver side.

For simplicity, an odd number of \(2K_\text{h}+1\) horizontal elements and \(2K_\text{v}+1\) vertical elements are disposed on the surface of RIS; i.e. for a total of \(N = (2K_\text{h}+1)(2K_\text{v}+1)\) elements. 

In what follows, two configurations will be evaluated:
\begin{enumerate}
	\item A linear RIS with \(K_\text{h}=10\) and \(K_\text{v}=0\) for a total number of 21 \(\nicefrac{\lambda}{2}\)-spaced elements.
	\item A planar RIS with \(K_\text{h}=K_\text{v}=10\) for a total number of 441 \(\nicefrac{\lambda}{2}\)-spaced elements.
\end{enumerate}

\begin{figure}[b]
	\centerline{\includegraphics[width=0.5\textwidth]{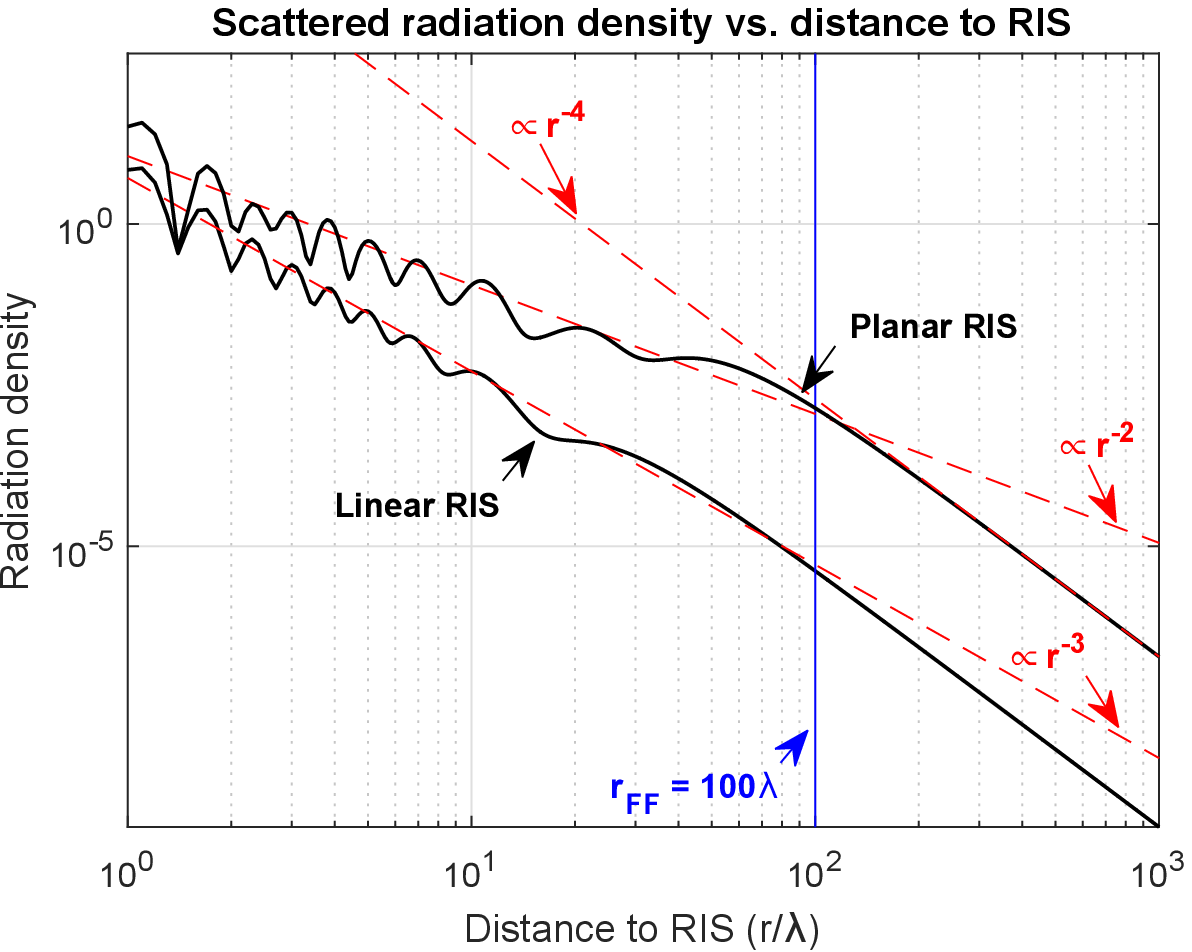}}
	\caption{Radiation density (power per unit area) of the field scattered by RIS versus distance for the setup of Fig.~\ref{fig:planarRIS}.}
	\label{fig:pathloss}
\end{figure}

As observed in Fig.~\ref{fig:pathloss}, the radiation density clearly exposes different behaviors for the array near-field (\(r<\rarray\)) and its far-field region (\(r>\rarray\)).

More specifically, the array far-field region unsurprisingly exposes a path-loss related to \(r^{-4}\) in both the linear and planar configurations. The operation over such a sector can be characterized through metrics used in the radar community (such as the radar cross section) as done in~\cite{WCNC2019}. Nonetheless, the latter is clearly not the most interesting region of operation for the RIS.

On the other hand, the array near-field exposes a seemingly oscillatory behavior around \(r^{-2}\) and \(r^{-3}\) for the planar and linear configurations, respectively. The latter is explained by the fact that, through its finite number of antenna elements, the RIS is sampling the field at discrete points in space. 

In particular, the oscillations illustrate the constructive and destructive interference caused by the complex exponential terms in~\eqref{eq:RDISUncIso} as induced by the Green function. Nonetheless, as it is shown in section~\ref{s:RISBTFS}, such oscillations can be compensated for through smart dephasing.

\subsection{The Fresnel zone perspective}
\label{s:fresnel}
Previously, it was shown that the operation of a planar RIS in its near-field uncovers a behavior that resembles free-space propagation, in particular, under a non-line of sight transmitter-receiver link. In spite of this, the derived mathematical model does not immediately offer an insightful understanding from the perspective of the propagating wave.

Thus, the aim of this section is to develop an intuition on how this is possible and, more specifically, on the interaction of the size of RIS and its near-field region in relation to the Fresnel zones at the RIS interface. 

As a preamble, such an intuition will be confirmed in section~\ref{s:validation} through a simple yet insightful simulation on a commercial electromagnetic solver\footnote{-- i.e. WIPL-D, \textit{https://wipl-d.com/}}, based on the well known method of moments (MoM). In addition, the latter will not only corroborate the predictions of section~\ref{s:RISQff}, but it will open the possibility for a path-loss better than the free-space one, as revealed in section~\ref{s:RISBTFS}.

\begin{figure}[t]
	\centerline{\includegraphics[width=0.5\textwidth]{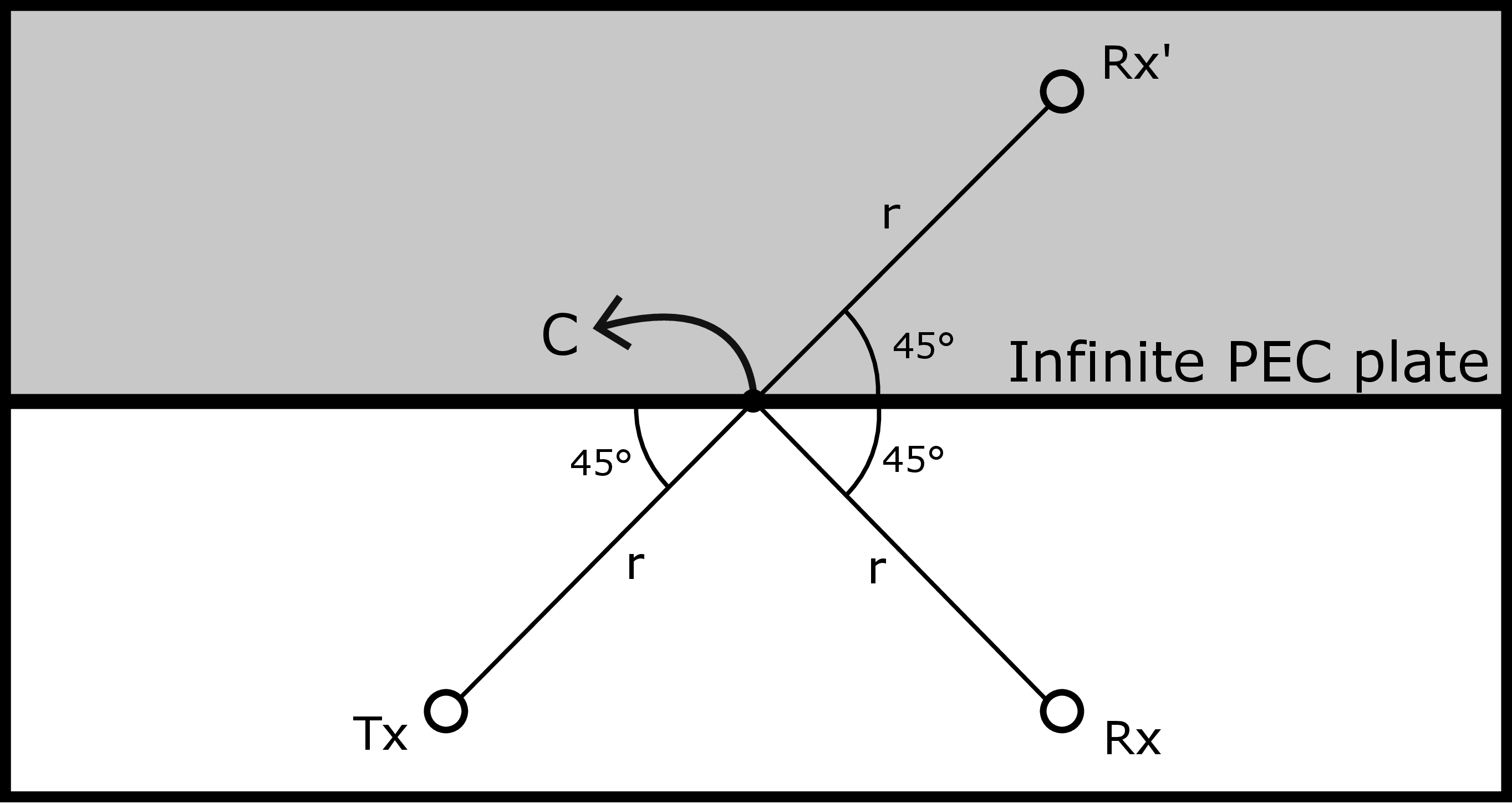}}
	\caption{Schematic view of the Tx, Rx for mirror reflection.}
	\label{fig:reflect}
\end{figure}

Let us begin by considering Fig.~\ref{fig:reflect}, where an unobstructed transmitter-receiver link  is presented. Specifically, note the presence of an infinite perfect electrically conducting (PEC) plane parallel to the line joining the transmitter and receiver sides.

Note that, by virtue of the well-known principle of images, the PEC plane can be removed to study separately the line-of-sight and the reflection. 

\begin{figure}[t]
	\centerline{\includegraphics[width=0.45\textwidth]{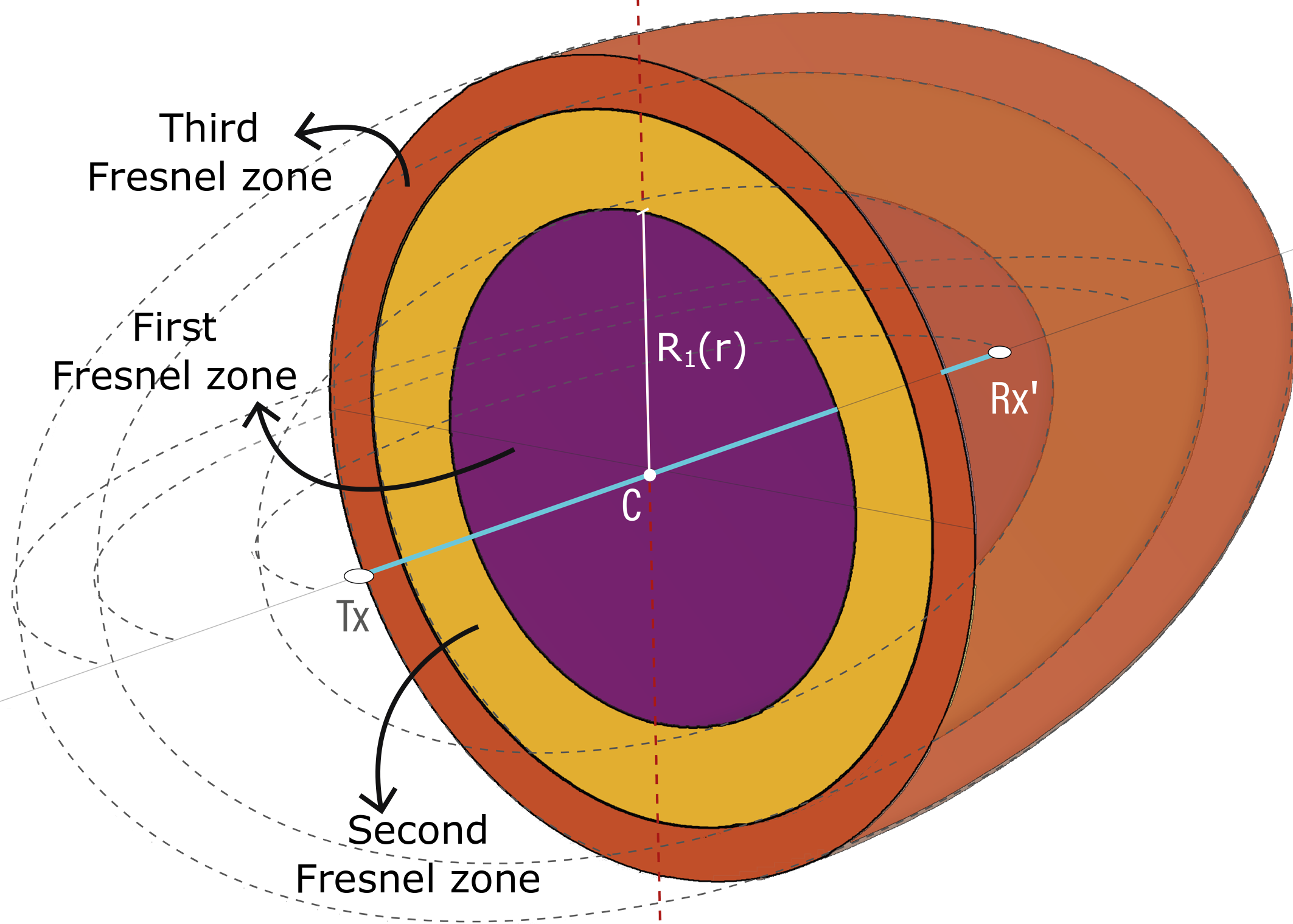}}
	\caption{View of the equivalent Tx-Rx' scenario for the contribution of the reflector.}
	\label{fig:ellipse}
\end{figure}

In fact, the contribution of the reflection is obtained by mirroring the receiver side and studying the equivalent environment. Thus, the equivalent setup of Fig.~\ref{fig:ellipse} is used in what follows to uncover the spatial distribution of the fields around such an interfacing plane.

More specifically, Fig.~\ref{fig:ellipse} shows the transmitter as well as the mirrored image of the receiver in a perfectly unobstructed environment. Additionally, the first three Fresnel zones are presented as a way to understand the most contributing regions.

The \(l\)\textsuperscript{th} Fresnel zone is a region of space whose boundaries are ellipsoids defined as the paths with \((l{-}1)\,\pi\) and \(l\,\pi\) propagation phase-shifts with respect to the shortest central path~\cite{FRESNEL} for its inner and outer boundaries, respectively. Additionally, the transmitter and receiver locations are the focal points of such ellipsoids, whose boundaries at the midpoint C can be approximated by:
\begin{equation}
\label{eq:Fresnel}
	R_{l}(r)\approxeq\sqrt{\frac{l\lambda}{2}\,r\,},\,\,r\gg l\lambda,
\end{equation}
where \(R_{l}(r)\) corresponds to the radius of the l\textsuperscript{th} Fresnel zone for the symmetrical (in \(r\)) arrangement of Fig.~\ref{fig:reflect}.

\begin{figure}[b]
	\centerline{\includegraphics[width=0.4\textwidth]{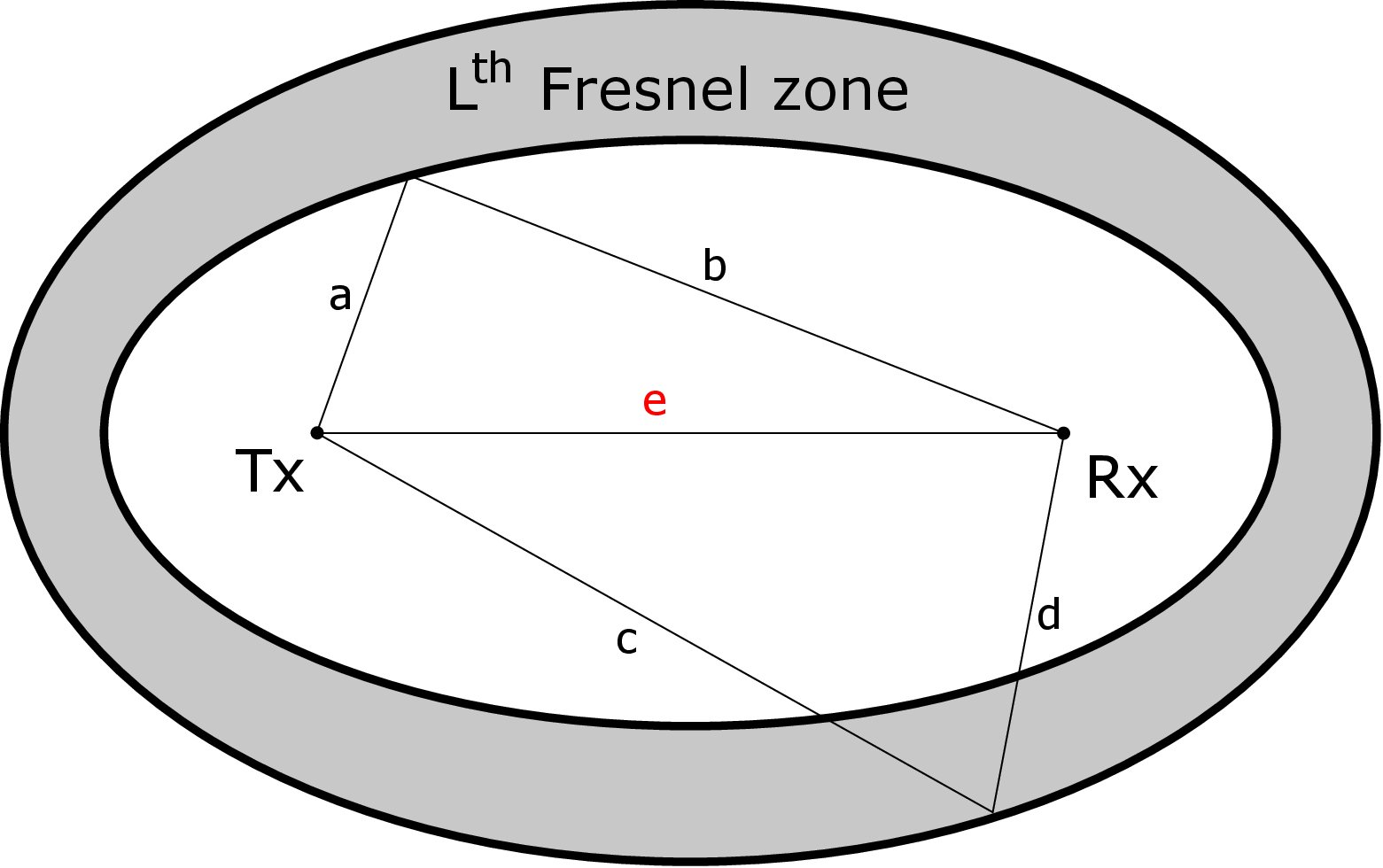}}
	\caption{Geometrical representation of the l\textsuperscript{th} Fresnel zone.}
	\label{fig:FresnelZone}
\end{figure}

It is well known that the contribution of the first Fresnel zone is the most important one. In order to understand why that is the case, note the following:
\begin{enumerate}[label=\arabic*),ref=\arabic*)]
	\item As shown in Fig.~\ref{fig:FresnelZone}, the Fresnel zones are defined as ellipsoids with constant propagation phase \textbf{relative} to the shortest path between the transmitter and receiver. As a consequence, the closer the transmitter and receiver sides, the smaller the Fresnel zones and vice-versa.
	\label{it:1}
	\item The Fresnel zone boundaries get closer for increasing \(l\):
	\begin{equation*}
		R_{l+1}(r) - R_{l}(r)= \sqrt{r\frac{\lambda}{2}}\,\mathcal{O}\Big(\sqrt{l\,}\,\Big)
	\end{equation*}
	\label{it:2}
	\item There is a phase difference of \(2\pi\) between any pair of paths distanced two Fresnel zones from each other.
	\label{it:3}
\end{enumerate}

Therefore, in the extreme of large \(r\), high order Fresnel zones (i.e. higher than one) are significantly weak relative to the first zone and, thus, they do not contribute significantly to the received power; see \ref{it:1}. 

On the contrary, for smaller values of \(r\), high order Fresnel zones are almost equally strong due to~\ref{it:2} but, at the same time, they interfere destructively with their successive one as a result of~\ref{it:3}.

\begin{figure}[t]
	\centerline{\includegraphics[width=0.5\textwidth]{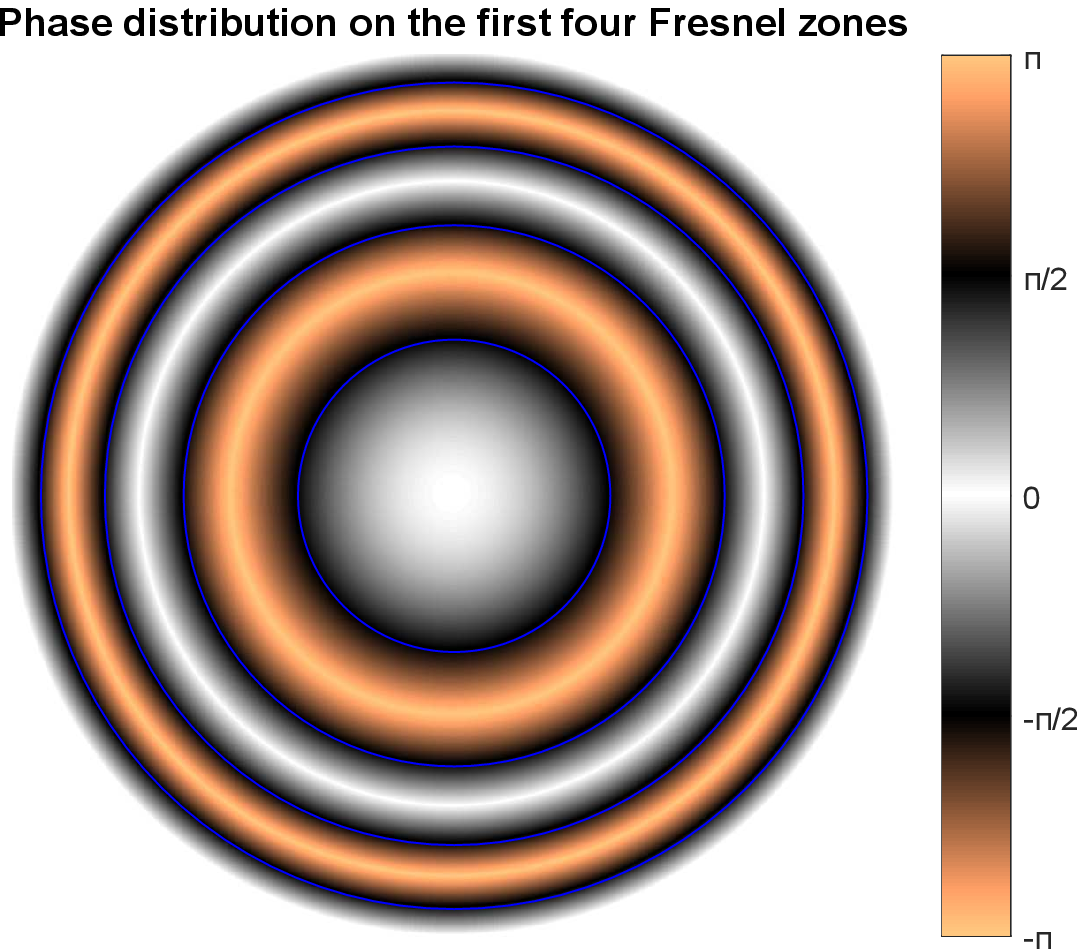}}
	\caption{Phase distribution on the first four Fresnel zones. White and orange represent constructive and destructive interference, respectively.}
	\label{fig:FresnelPhase}
\end{figure}

In particular, note that even numbered Fresnel zones \textbf{always} interfere destructively; as opposed to odd numbered ones that interfere constructively as shown in Fig.~\ref{fig:FresnelPhase}. The reader might realize at this point how, by a proper dephasing, RIS could in principle outperform free-space propagation.

To continue, it might be useful to look at an infinitely large mirror as spatially integrating the fields over such an infinite aperture. As explained by the equivalence principle over the scenario of Fig.~\ref{fig:reflect}, such a spatial integration converges to the \(r^{-2}\) path-loss dependence we are so familiarized to. 

In fact, if the mirror were finite and centered in the shortest transmitter-receiver path, the spatial integration would be truncated. At the same time, due to the Fresnel zone resizing, such a truncation would expose oscillations if the transmitter and receiver were symmetrically moved. 

The latter gives an intuitive understanding to the array near-field behavior of Fig.~\ref{fig:pathloss}.

Finally, as a mean of linking RIS' field decomposition and the Fresnel zone perspective, observe from~\eqref{eq:Fresnel} (as well as from~\eqref{eq:rgloballocal} in Appendix~\ref{ap:EJqff}) that \(R_{l=1}(\rarray) = D\); with \(\rarray\) being the low limit of the array far-field and \(D\) being the \textit{visible} dimension\footnote{By visible dimension we refer to the smallest diameter of a circle located on the plane transversal to the direction of propagation and containing the array.} of the array. 

In other words, the RIS is being operated in its near-field region when at least the first Fresnel zone of the transmitter-receiver equivalent path (see Fig.~\ref{fig:ellipse} and Fig.~\ref{fig:FresnelPhase}) is perfectly contained within the RIS itself.

\subsection{Model and intuition corroboration}
\label{s:validation}

In order to confirm that the last assertions are indeed correct, and that the model of section~\ref{s:RISQff} predicts the right phenomena, we have evaluated a very simple scenario using WIPL-D. 

More specifically, a two port setup with two vertically-polarized half-wave dipoles (acting as the transmitter and receiver) in the presence of a finite metal plate was simulated. The size of the plate was fixed to \(10\lambda\times10\lambda\) and the dipoles were positioned symmetrically (\(45^\circ\) from the vector normal to the plate's surface) a distance \(r\); exactly like for the RIS setup of section~\ref{s:RISQff}. 

In particular, the power transmission coefficient \(|S_{2,1}|^2\) was computed as we are technically operating over the near-field of the plate (i.e. making far-field metrics such as the radar cross section invalid). 

Additionally, as the aforementioned setup computes the net (direct+reflected) fields, separate simulations (with and without the plate) were done in order to subtract the the direct path and obtain the reflected contribution.

\begin{figure}[t]
	\centerline{\includegraphics[width=0.5\textwidth]{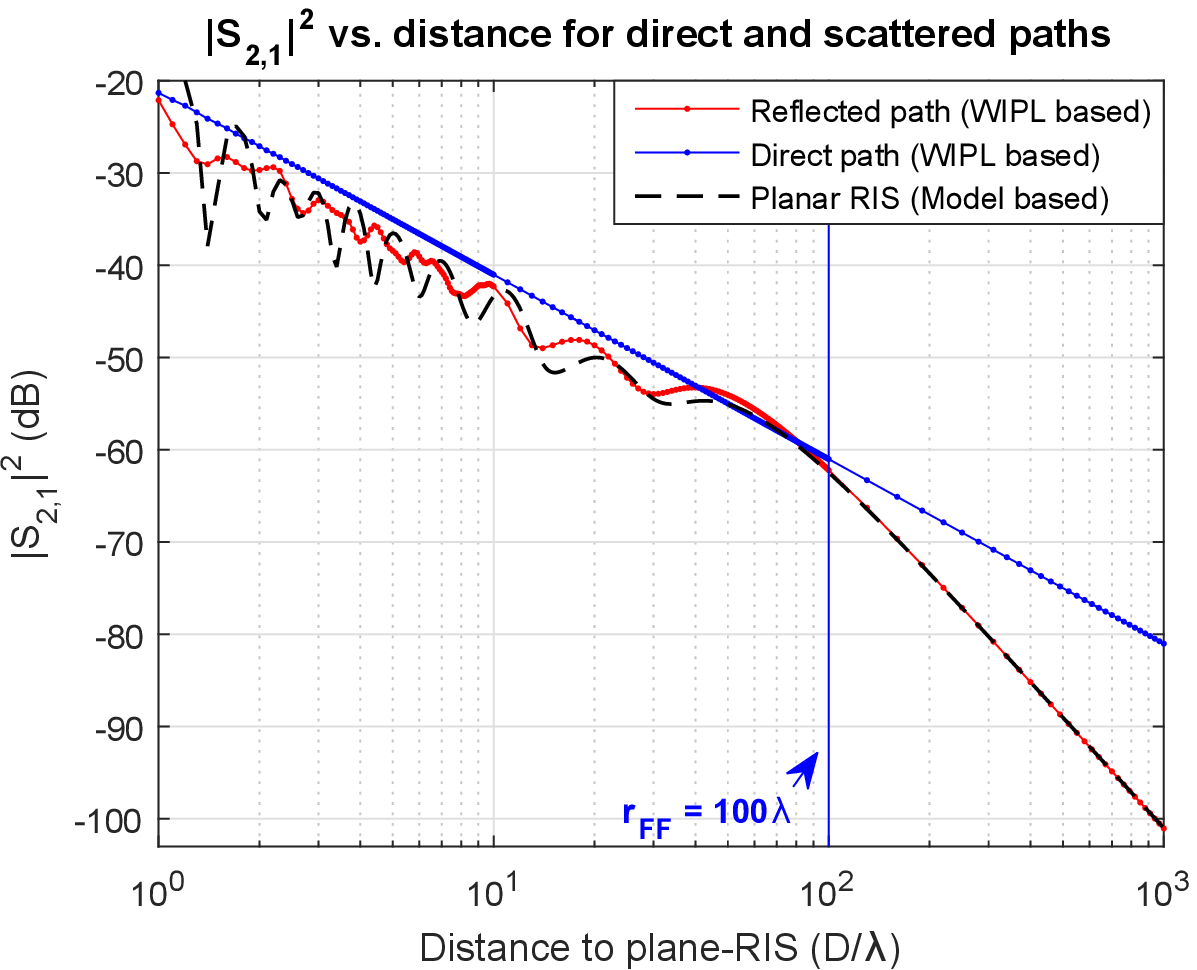}}
	\caption{\(|S_{2,1}|^2\) quantifying the power received through reflection from a finite metal plate vs. transmitter-receiver distance.}
	\label{fig:WIPL}
\end{figure}

As observed in Fig.~\ref{fig:WIPL}, the reflected path exhibits the expected oscillatory behavior at the near-field region of the plate. Note as well that, in spite of the discrete nature of the RIS, the model is able to capture with fair accuracy the details of its continuous-equivalent\footnote{Naturally, RIS' curve was vertically shifted to make it coincide with WIPL's results in the array far-field as, in particular, the multiplicative coefficients in~\eqref{eq:RDISUncIso} cannot be determined for hypothetical isotropic antennas.}. Nonetheless, it can be shown that the WIPL numerical results and the model converge when the element density of the model is increased within such a confined region of space.
\newpage
\section{Free-space propagation\\outperformed through RIS}
\label{s:RISBTFS}

Recall that, from section~\ref{s:fresnel}, the contribution of the even numbered Fresnel zones is always destructive. As a consequence, if such zones are contained within the RIS\footnote{Namely, if it is being operated in the array near-field region.}, a dense enough architecture might in principle be able to compensate for their destructive nature.

As a matter of fact, in connection with the notion established in section~\ref{s:fresnel}, the model in section~\ref{s:RISQff} allows to transparently predict the maximum obtainable power for an architecture-specific RIS. 

In particular, a smart-enough RIS would, at its best, compensate for the path-related phase shift; giving for the received radiation density:
\begin{equation}
\label{eq:RDISUncIsoNP}
\Ps_\textrm{max}(\vec{r}_\rx,\vec{r}_\tx) = \frac{k^4\eta^3}{2|Z_A|^2}\big|F_{\theta}^\textrm{iso}\big|^6 |I_\tx|^2 \Bigg|\sum\limits_{n=1}^N \frac{1}{4\pi r_{\textrm{r},n}}\,\frac{1}{4\pi r_{\textrm{t},n}}\Bigg|^2,
\end{equation}
using the notation of~\eqref{eq:RDISUncIso} to represent all involved quantities.

Such a radiation density for the architecture of section~\ref{s:RISQff} is shown and compared to the short-circuit, i.e. \(\vec{x}=\vec{0}\,\Omega\), in the following.

\begin{figure}[t]
	\centerline{\includegraphics[width=0.5\textwidth]{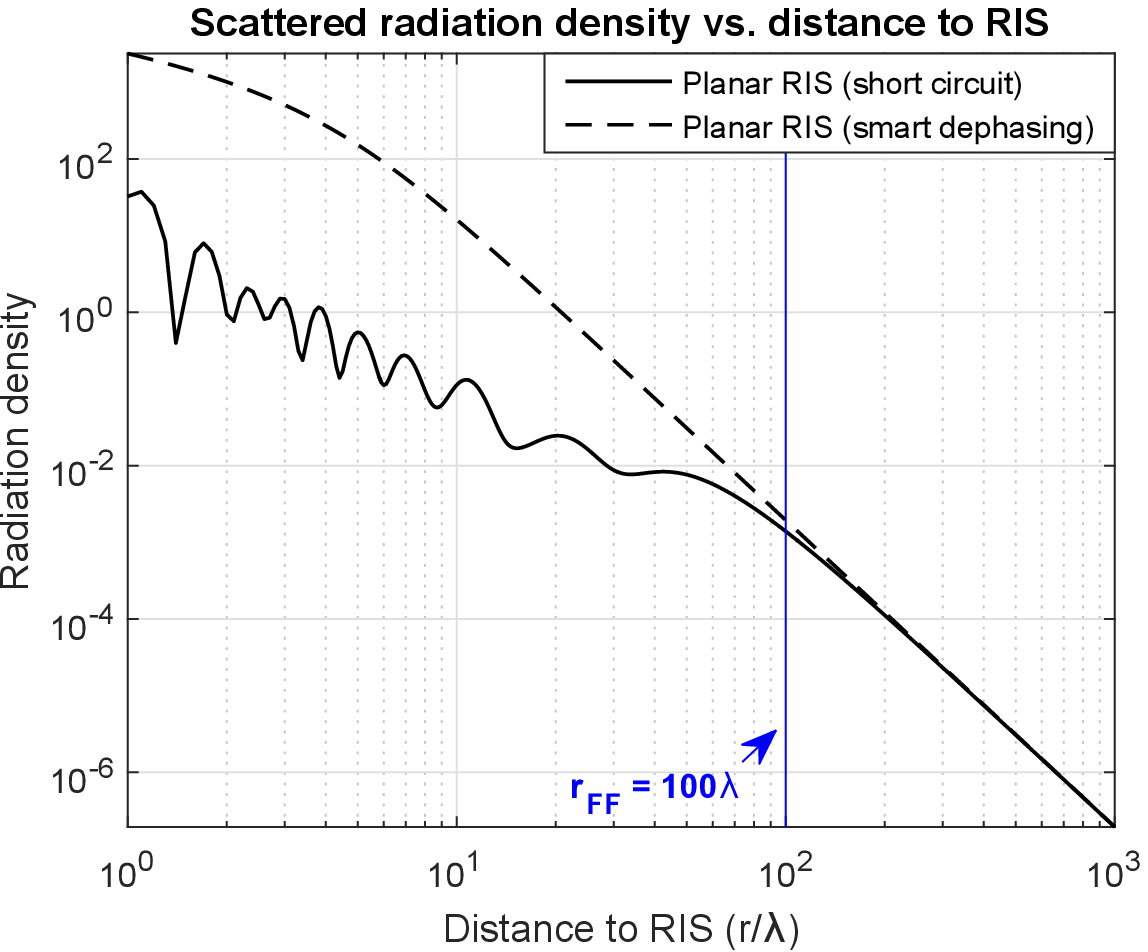}}
	\caption{Radiation density (power per unit area) of the field scattered by the RIS with smart dephasing versus distance.}
	\label{fig:smartRIS}
\end{figure}

As observed in Fig.~\ref{fig:smartRIS}, the path-loss for the mentioned smart-dephasing technique does deviate from the conventional \(r^{-2}\) in the array near-field region. In fact, it exposes a behavior that outperforms free-space propagation for a completely-obstructed (NLOS) transmitter-receiver link. 

Note, as well, that the lack of oscillations in the dashed curve is a direct consequence of the removal of the (even-numbered) Fresnel zone destructive nature; as expected from section~\ref{s:fresnel}. 

It must be stressed that an element spacing of \(\nicefrac{\lambda}{2}\) was enough in Fig.~\ref{fig:smartRIS} to maintain a constant log-log increase of receive power with decreasing distance\footnote{-- to be specific, of 40 dB per decade.} for a significant part of the array near-field region. 

More importantly, the behavior observed in Fig.~\ref{fig:smartRIS} implies that the dependence with the fourth power of the distance is a consequence of the constant phase along the RIS. Typically, the latter manifests at the array far-field region where, as explained in section~\ref{s:fresnel}, the RIS is fully contained within the first Fresnel zone. Nonetheless, as explored in this section, such a behavior can be artificially enforced in the array near-field.

In particular, while a \(r^{-4}\) dependence at large distance may be seen as poor, it turns out to be advantageous when approaching the mirror below \(\rarray\) as a way to avoid the transition to the (inferior) \(r^{-2}\) regime.

\section{Complete link-level system characterization}
\label{s:llmodel}
To finish up, the authors would to like briefly summarize the implications of section~\ref{s:prop} to the familiar link-level characterization of communication systems. 

For simplicity, in what follows, we will assume single-element, single-polarization and isotropic transmitter and receiver sides. On the other hand, the RIS elementary unit can be arbitrarily defined as its impact is accounted for in the generalized array manifold of~\eqref{eq:AMv}. 

It must also be added that the multi-antenna transmitter/receiver extension can be straightforwardly envisaged by virtue of the superposition principle. Nonetheless, its mathematical representation can easily become cumbersome as a result of the multi-location dependencies\footnote{-- in particular for the case in which the array elementary units of the transmitter and receiver sides are not isotropic anymore.}.

Therefore, relying on the RIS model derivation of section~\ref{s:RISQff}, the complete link-level system model can be shown equivalent to:
\begin{equation}
\label{eq:linkModel}
	y = \Big(\tilde{h}_\textrm{tr}\,G(r_\textrm{tr})+\tilde{h}_\textrm{RIS}\,\vec{a}_p^H(\vec{r}_\rx)\big(\vec{Z}+\j\vec{X}_L\big)^{-1}\,\vec{a}_p(\vec{r}_\tx)\Big)\,s+n,
\end{equation}
where \(s\), \(y\) and \(n\) are the conventional input, output and additive white Gaussian noise at the receiver side, respectively; \(G(r)\) is the free-space Green function of Appendix~\ref{ap:EJ}; \(r_\textrm{tr}\) is the shortest-path transmitter-receiver distance; \(\vec{r}_\tx\) and \(\vec{r}_\rx\) are the transmitter and receiver locations relative to RIS' coordinate reference, respectively; and \(\vec{a}_p(\vec{r})\in\C^N\) is the generalized array manifold of~\eqref{eq:AMv} (see Appendix~\ref{ap:EJqff} for its derivation) polarized along \(p\) for an \(N\) element RIS.

Additionally, in~\eqref{eq:linkModel}, \(\tilde{h}_\textrm{tr}\) and \(\tilde{h}_\textrm{RIS}\) are spatially-flat channel coefficients that represent a scenario in which all links (transmitter-receiver, transmitter-RIS and RIS-receiver) are dominated by their line-of-sight components. These channel coefficients also absorb all physical quantities that are not of concern for link-level characterization; allowing to introduce the dimensionless \textit{signal} denomination\footnote{-- where a signal is simply defined as an observable change in a quantity.}. 

Note that, if the transmitter-receiver link is either obstructed or suffers from strong multi-path propagation, its impact shall be embedded onto \(\tilde{h}_\textrm{tr}\). On the other hand, we do not expect \(\tilde{h}_\textrm{RIS}\) to be greatly impacted by multi-path propagation as higher order scattering might strongly attenuate contributions other than the line-of-sight ones.

\section{Conclusion}
In this work, we have presented a view that unifies the opposite behavior of RIS as a scatterer and as a mirror. In particular, relying on fundamental electromagnetics, we do so as a mean to identify scenarios and show the strong potential behind the RIS concept.

We have shown that, depending on its size and distance, the RIS can be observed as a zero, one or two dimensional object whose radiated power exposes a dependence with the fourth, third or second power of the distance, respectively.

Additionally, we have employed the Fresnel zone decomposition to build an intuitive understanding of the interplay of the involved quantities. More specifically, we have uncovered the role of the phase in determining the ultimate path-loss exponent and to show how, through smart dephasing, free-space propagation can be outperformed.

Moreover, as one of the main contributions of the current work, we have presented a model for the signal-level characterization of a transmitter-receiver link in the presence of RIS. Particularly, such a model concisely captures all described phenomena over all the regions of operation.

To finalize, even though the amount of elements studied in this work might seem tremendous, it must be kept in mind that the described planar architecture would barely occupy \(1\,\text{m}^2\) at a central frequency of 3 GHz.
More importantly, it would present a path-loss more favorable than free-space (under a completely obstructed direct transmitter-receiver link) for a distance up to \(\rarray=10\,\text{m}\), on such a frequency of operation.


%

\appendices
\section{The far-field radiated by an antenna}
\label{ap:EJ}
Recall that the electric field intensity radiated by a source current density is given in terms of the so-called a magnetic potential vector \(\vec{A}\)~\cite{ewa}:
\begin{equation}
\label{eq:EA}
\vec{E} = \frac{1}{\j\omega\mu\epsilon}\Big[\nabla\big(\nabla\cdot\vec{A}\big)+k^2\vec{A}\Big],
\end{equation}
where \(\epsilon\) and \(\mu\) are the electric permittivity and magnetic permeability of the propagation medium, and \(\omega=2\pi f\). 

Particularly, the magnetic potential vector (used as a convenient step to obtain \(\vec{E}\)) is obtained as the convolution of the source current density \(\vec{J}(\vec{r})\) with the free-space Green function of the Helmholtz equation~\cite{ewa}:
\begin{equation}
\label{eq:GF}
\nabla^2 G(\vec{r}) + k^2 G(\vec{r}) = -\delta(\vec{r}),\,\,\,G(\vec{r})\defeqto\frac{\e^{-\j k|\vec{r}|}}{4\pi |\vec{r}|},
\end{equation}
where \(G(\vec{r})\) is the Green function, \(k=\nicefrac{2\pi}{\lambda}\) is the wave-number and \(\delta(\vec{r})\) is the three-dimensional delta function.

In mathematical form, the magnetic potential vector reads:
\begin{equation}
\label{eq:Aint}
\vec{A}(\vec{r}) = \cur \mu \int_{V^\prime} \vec{J}(\vec{r}^\prime)\,G(\vec{r}-\vec{r}^\prime)\,d^3\vec{r}^\prime,
\end{equation}
where \(\cur\) is the input current, \(\vec{J}(\vec{r})\) is the current distribution of the element normalized to such an input current, \(\vec{r}\) is the field (observation) point and \(\vec{r}^\prime\) is the source (integration) point (i.e. over \(V^\prime\) that is a volume containing all sources). 

At this point, we ought to highlight the linearity of the integro-differential operator in~\eqref{eq:Aint} with the input current that propagates until \(E\) as per~\eqref{eq:EA}. 

In particular, it is because of such a linearity that we are allowed to resort to tools such as the array factor as a mean of describing the behavior of arrays in terms of their elementary unit. 

\subsection{Far-field approximation}

The shifted argument of the Green function, i.e. \(G(\vec{r}-\vec{r}^\prime)\) in~\eqref{eq:Aint}, can be approximated for the region known as far-field through:
\begin{equation}
\label{eq:ffapp}
G(\vec{r}-\vec{r}^\prime) \approxeq \frac{\e^{-\j k\,(|\vec{r}|-\vec{\hat{r}}\cdot\vec{r}^\prime)}}{4\pi |\vec{r}|}\text{, for }|\vec{r}|\gg l\text{ and }|\vec{r}|\gg \frac{2l^2}{\lambda},
\end{equation}
where \(l\) is the largest dimension of the smallest integration volume \(V^\prime\) in~\eqref{eq:Aint} containing all sources and \(\vec{\hat{r}}\) is a unitary vector pointing at the far-field observation point. 

Therefore, in~\eqref{eq:ffapp}, the dependence with the shifted observation point in the numerator was replaced by a first order approximation whereas, in the denominator, it was replaced by an approximation of order zero.

Moreover, solving~\eqref{eq:Aint} with~\eqref{eq:ffapp} and plugging it in~\eqref{eq:EA}, the radiated electric field intensity \(\vec{E}\) can be shown equal to~\cite{ewa}:
\begin{equation}
\label{eq:EF}
	\vec{E}(\vec{r})=-\j \cur k\eta\,\frac{\e^{-\j kr}}{4\pi r}\,\vec{F}_\perp(\vec{\hat{r}}),\,\,\vec{F}_\perp(\vec{\hat{r}})=\vec{F}(\vec{\hat{r}})\times\vec{\hat{r}},
\end{equation}
where \(\vec{F}(\vec{\hat{r}})\) is a far-field measure known as the radiation vector and is explicitly defined as: 
\begin{equation}
\label{eq:F}
\vec{F}(\vec{\hat{r}}) \defeqto \int_{V^\prime}\vec{J}(\vec{r}^{\prime})\,\e^{\j k\vec{\hat{r}}\cdot\vec{r}^{\prime}}\,d^3\vec{r}^{\prime},
\end{equation}
with \(V^\prime\) fully containing the source current distribution.

To conclude, it must be noted that \(\vec{F}_\perp(\vec{\hat{r}})=\vec{F}(\vec{\hat{r}})\times\vec{\hat{r}}\) in~\eqref{eq:EF} is also known as the effective length (vector) of the antenna. In particular, the effective length vector is of relevance when studying far-field radiation incident to the antenna.

More specifically, the electro-motive force \(\vo\) at the antenna terminals can be written simply as~\cite{BOOKBALANIS}:
\begin{equation}
	\vo = \vec{F}_\perp(\vec{\hat{r}})\cdot\vec{E}_\inc,
\end{equation}
where \(\vec{E}_\inc\) is the electric field intensity characterizing the incident radiation at the location of the receiving antenna under consideration.

\section{The array near-field radiation}
\label{ap:EJqff}
Let us continue by considering the problem of determining the electric field intensity \(\vec{E}\) at the array near-field region resulting from a multi-antenna arrangement as source of fields. 

In particular, as it will be clear in what follows, the characterization at the array near-field region captures the behavior over the array far-field as a particular case. Thus, allowing to model RIS on both such regions of interest.

Therefore, as a first step, consider the source current density of a multiple antenna architecture such as:
\begin{equation}
\label{eq:linsup}
\vec{J}(\vec{r})=\sum\limits_{n=1}^N\cur_n\,\vec{J}_0(\vec{r}-\vec{r}_n),
\end{equation}
where \(N\) is the number of elements, \(\vec{r}_n\) is the location of the n\textsuperscript{th} element with respect to a common reference, \(\cur_n\) is the input current at the n\textsuperscript{th} element and \(\vec{J}_0(\vec{r})\) is the (identical) current distribution of the array elementary unit normalized to such an input current. 

Based on Appendix~\ref{ap:EJ}, the magnetic potential vector for such an architecture can be written as\footnote{-- observe that we don't resort directly to the radiation vector as this one would inherently solve the problem in the array far-field region.}:
\begin{equation}
\label{eq:Aint1}
\vec{A}(\vec{r})= \mu\int_{V^\prime} \sum\limits_{n=1}^N\cur_n\,\vec{J}_0(\vec{r}^\prime-\vec{r}_n)\,G(\vec{r}-\vec{r}^\prime)\,d^3\vec{r}^\prime,
\end{equation}
where \(V^\prime\) should include all the sources represented by~\eqref{eq:linsup}. 

Recall that the most commonly used antenna metrics (directivity, gain, antenna aperture, etc.) give an approximately correct characterization for the far-field region of the antenna or antenna array under consideration. 


Nonetheless, note that the lower limit of the array and element far-field regions are given in terms of the largest dimension of the array \(D\) and its elementary unit \(D_0\) by:
\begin{equation}
\label{eq:rgloballocal}
\rarray=\frac{2D^2}{\lambda},\,\,\relem=\frac{2D_0^2}{\lambda},
\end{equation}
where \(\lambda=\nicefrac{c}{f}\) corresponds to the wavelength of operation and \(c\) to the speed of light.

As a consequence, the far-field approximation of the Green function of~\eqref{eq:ffapp} cannot be used in~\eqref{eq:Aint1} to compute \(\vec{A}\) over the array near-field region of interest (i.e. \(\relem<|\vec{r}|<\rarray\)). 

On the other hand, in the following, we will resort to a different strategy as a mean of approximating it at the region of interest.


\subsubsection*{The far-field condition revisited}
if the integration and summation are swapped in~\eqref{eq:Aint1}, such an expression can be rewritten as:
\begin{equation}
\label{eq:Aint2}
\vec{A}(\vec{r})= \mu\sum\limits_{n=1}^N\cur_n\int_{V^{\prime\prime}}\vec{J}_0(\vec{r}^{\prime\prime})\,G\big((\vec{r}-\vec{r}_n)-\vec{r}^{\prime\prime}\big)\,d^3\vec{r}^{\prime\prime},
\end{equation}
where the substitution \(\vec{r}^{\prime\prime}=\vec{r}^\prime-\vec{r}_n\) was used and accounted for in the volume of integration.

Note that, if \(\vec{J}_0(\vec{r})\) is concentrated in a closed domain over \(\vec{r}\in\R^3\) and \(|\vec{r}_n-\vec{r}_m|> d_\text{max}\,\forall n\neq m\) with \(d_\text{max}\) being the largest dimension of such a closed domain,~\eqref{eq:Aint2} can be expressed as:
\begin{equation}
\label{eq:Aint3}
\vec{A}(\vec{r})= \mu\sum\limits_{n=1}^N\cur_n\int_{V^{\prime\prime}_n}\vec{J}_0(\vec{r}^{\prime\prime})\,G\big((\vec{r}-\vec{r}_n)-\vec{r}^{\prime\prime}\big)\,d^3\vec{r}^{\prime\prime},
\end{equation}
where \(V^{\prime\prime}=\bigcup_{n=1}^N V^{\prime\prime}_n\) with \(V_n^{\prime\prime}\) tightly enclosing the domain over which \(\vec{J}_0(\vec{r}-\vec{r}_n)\) is concentrated and, more importantly, such regions are disjoint, i.e. \(V_i^{\prime\prime}\bigcap V_j^{\prime\prime}=\varnothing\,\,\forall\,i\neq j\).

The importance of the previous result lies on that, while the far-field Green function cannot be used in~\eqref{eq:Aint2}, it can be used over the separate domains of integration in~\eqref{eq:Aint3}. The latter, as long as \(|\vec{r}-\vec{r}_n|>\relem\,\forall n\) with \(\relem\) given by~\eqref{eq:rgloballocal}.

In particular, the expression~\eqref{eq:Aint3} can be largely simplified by identifying the radiation vector of~\eqref{eq:F} through:
\begin{equation}
\label{eq:Aint4}
\vec{A}(\vec{r})= \sum\limits_{n=1}^N\cur_n\overbrace{\mu\underbrace{\frac{\e^{-\j k|\vec{r}-\vec{r}_n|}}{4\pi |\vec{r}-\vec{r}_n|}}_{G(\vec{r}-\vec{r}_n)}\vec{F}_0\bigg(\frac{\vec{r}-\vec{r}_n}{|\vec{r}-\vec{r}_n|}\bigg)}^{\vec{A}_0^\textsc{ff}(\vec{r}-\vec{r}_n)},
\end{equation}
where \(\vec{A}_0^\textsc{ff}(\vec{r})\) is identified as the far-field approximation of the magnetic potential vector of the array elementary unit, \(G(\vec{r})\) as the Green function in~\eqref{eq:GF} and \(\vec{F}_0(\vec{\hat{r}})\) is the radiation vector of the elementary array unit of Appendix~\ref{ap:EJ}.

Note in~\eqref{eq:Aint4} that \(\vec{F}_0(\vec{\hat{r}})\) depends \textbf{exclusively} on the direction of the observation point relative to the location of the n\textsuperscript{th} element\footnote{-- as this last one is a far-field measure with respect to such an element.}. Moreover, unlike the conventional array far-field, the radiation vector cannot be factored out of the summation and, therefore, an array factor cannot be defined anymore.

By properties of the operators in~\eqref{eq:EA}, given that the argument of \(\vec{A}_0^\textsc{ff}(\vec{r})\) in~\eqref{eq:Aint4} is simply translated on every summation term, the total radiated field in the array near-field region can be written as:
\begin{equation}
\label{eq:ET}
\vec{E}(\vec{r})=-\j k\eta\sum\limits_{n=1}^N\cur_n\,G(\vec{r}-\vec{r}_n)\,\vec{F}_{0,\perp}\bigg(\frac{\vec{r}-\vec{r}_n}{|\vec{r}-\vec{r}_n|}\bigg).
\end{equation}

Note also that, for the general case of dual polarized transmitting antennas, such a total radiated field can be written in terms of its p polarization as:
\begin{equation}
E_p(\vec{r})=-\j k\eta\sum\limits_{n=1}^N\cur_n\,a_{n,p}(\vec{r}).
\end{equation}
where \(a_{n,p}(\vec{r})\) is an order-2 tensor quantity called here the generalized array manifold; formally defined as:
\begin{equation}
\label{eq:AM}
a_{n,p}(\vec{r})\defeqto G(\vec{r}-\vec{r}_n)\,F_{0,p}\bigg(\frac{\vec{r}-\vec{r}_n}{|\vec{r}-\vec{r}_n|}\bigg)\,\forall\,n\le N,
\end{equation}
with \(G(\vec{r}-\vec{r}_n)\) being the translation of the Green function of~\eqref{eq:GF} and \(F_{0,p}(\vec{r})\) denoting the radiation vector along the p direction of polarization\footnote{note that \(\vec{\hat{p}}\) must always be orthogonal to \(\vec{\hat{r}}\).}.

Particularly, observe that if single p-polarized radiation is considered, the array manifold in~\eqref{eq:AM} collapses into a vector that can be simply denoted as \(\vec{a}_p(\vec{r})\big|_n = a_{n,p}\,\forall\,n\le N\); for N array elements. 

Moreover, if~\eqref{eq:ET} is to be evaluated in the array far-field region (i.e. \(|\vec{r}|>\rarray\)), the conventional array factor can be recovered by replacing the Green function with its far-field approximation of~\eqref{eq:ffapp}. Thus, showing that~\eqref{eq:AM} indeed generalizes the array manifold with the array far-field region as a special case.
%
%
%
%
%

\section*{Note to the reader}
Observe that the definition of the radiation vector in terms of the \textbf{normalized} current density makes the radiation vector be slightly different to the one in~\cite{ewa}. 

As a matter of fact, their definition is equal to ours multiplied by the input current. Nonetheless, this makes the link between transmission and reception modes straightforward; i.e. through the same metric (not having to introduce a different notation for the effective length vector as these become essentially equivalent).

\ifCLASSOPTIONcaptionsoff
  \newpage
\fi

\end{document}